\documentclass[12pt,a4paper,final]{article} 
\usepackage{times}
\usepackage{a4wide}
\usepackage{amsfonts}
\usepackage{amssymb}
\usepackage{amsmath}
\usepackage{bbm}
\usepackage{booktabs} 
\usepackage[notref,notcite]{showkeys}
\usepackage{ifpdf}
\ifpdf
\usepackage[pdftex,unicode,implicit]{hyperref}
\hypersetup{%
  pdftitle    = {The Freudenthal gauge symmetry of the black holes of  
N=2,d=4 supergravity},
  pdfkeywords = {},
  pdfauthor   = {Pietro Galli, Patrick Meessen and Tomas Ortin},
  plainpages  = true,
  colorlinks  = true,
  citecolor   = blue,
  urlcolor    = red,
  linkcolor   = black
}
\newcommand{\hepth}[1]{{\tt
\href{http://www.arXiv.org/abs/hep-th/#1}{hep-th/#1}}}

\newcommand{\arxiv}[1]{{\tt
\href{http://www.arXiv.org/abs/#1}{#1}}}
\else
  \usepackage[dvips]{graphicx}
  \usepackage[unicode,implicit]{hyperref}
  \newcommand{\hepth}[1]{arXiv:{\tt hep-th/#1}}

  \newcommand{\arxiv}[1]{{\tt #1}}
\fi
\usepackage{tikz}\newcommand{\FPAUO}[2]{
\tikz[scale=.13,
         Uniovi/.style={color=green!51!blue, fill=green!51!blue}
 ] {
 \fill[Uniovi] (0,0) circle (10);
 \fill[white] (0,7) circle (1.5);
 \draw[Uniovi] (-2,7.5) rectangle (2,5.5);
 \fill[white] (-0.3,6.6) rectangle (0.3,0);   
 \fill[white] ( -0.9,6.2) rectangle (.9 ,5.6);
 \fill[white] (-1.4, 5.2) rectangle (1.4, 4.6);
 \fill[white] (0,0) ellipse (3.5 and 4);
 \fill[Uniovi] (-2.5,0.3) rectangle (2.5,-0.3);
 \fill[Uniovi] (-2,2.3) rectangle (2,1.7);
 \fill[Uniovi] (-2,-2.3) rectangle (2,-1.7);
 \fill[white] (-4.5,5.5) rectangle (-2.7,4.9);
 \fill[white] (-3.9,6.1) rectangle (-3.3,4.3);
 \fill[white] (4.5,5.5) rectangle (2.7,4.9);
 \fill[white] (3.9,6.1) rectangle (3.3,4.3);
 \foreach \x in { 0,..., 3 }
   \foreach \y in { 0,...,\x}
    {
     \fill[white] (-6-\x*0.7+\y*1.4,3.5-\x *1.97) -- (-5.6-\x*0.7+\y*1.4,2.4-\x *1.97) -- (-6.4-\x*0.7+\y*1.4,2.4-\x *1.97) -- cycle;
     \fill[white] (6-\x*0.7+\y*1.4,3.5-\x *1.97) -- (5.6-\x*0.7+\y*1.4,2.4-\x *1.97) -- (6.4-\x*0.7+\y*1.4,2.4-\x *1.97) -- cycle;
   };
 \draw (0,-6) node[
                               text centered, 
                               color=white, 
                               font={\fontsize{8}{4}\sffamily\selectfont}
                             ] {FPAUO-#1/#2};
}} 
\makeatletter
\@addtoreset{equation}{section}
\makeatother

\begin{document}

\begin{flushright}
\small
\FPAUO{12}{16}\\
IFIC/12-74\\
IFT-UAM/CSIC-12-94\\
November 30\textsuperscript{th}, 2012\\
\normalsize
\end{flushright}

\vspace{.5cm}

\begin{center}

{\Large {\bf The Freudenthal gauge symmetry of}}\\[.5cm] 
{\Large {\bf the black holes of $\mathcal{N}=2$, $d=4$ supergravity}}

\vspace{1cm}

\renewcommand{\thefootnote}{\alph{footnote}}
{\sl\large Pietro Galli$^{\clubsuit}$,}
\footnote{E-mail: {\tt Pietro.Galli [at] ific.uv.es}},
{\sl\large Patrick Meessen$^{\spadesuit}$}
\footnote{E-mail: {\tt meesenpatrick [at] uniovi.es}}
{\sl\large and Tom\'{a}s Ort\'{\i}n$^{\heartsuit}$}
\footnote{E-mail: {\tt Tomas.Ortin [at] csic.es}},
\renewcommand{\thefootnote}{\arabic{footnote}}

\vspace{1cm}

${}^{\clubsuit}${\it Departament de F\'{\i}sica Te\`orica and IFIC (CSIC-UVEG),
Universitat de Val\`encia,\\
C/ Dr.~Moliner, 50, 46100 Burjassot (Val\`encia), Spain}\\

\vspace{.2cm}

${}^{\spadesuit}${\it HEP Theory Group, Departamento de F\'{\i}sica, 
  Universidad de Oviedo\\ 
  Avda.~Calvo Sotelo s/n, 33007 Oviedo, Spain}\\

\vspace{.2cm}

${}^{\heartsuit}${\it Instituto de F\'{\i}sica Te\'orica UAM/CSIC\\
C/ Nicol\'as Cabrera, 13--15,  C.U.~Cantoblanco, 28049 Madrid, Spain}\\

\vspace{.5cm}


{\bf Abstract}

\end{center}

\begin{quotation}\small
We show that the representation of black-hole solutions in terms of the
variables $H^{M}$ which are harmonic functions in the supersymmetric case is
non-unique due to the existence of a local symmetry in the effective
action. This symmetry is a continuous (and local) generalization of the
discrete Freudenthal transformations initially introduced for the black-hole
charges and can be used to rewrite the physical fields of a solution in terms
of entirely different-looking functions. 
\end{quotation}

\newpage
\pagestyle{plain}


The FGK formalism developed in Ref.~\cite{Ferrara:1997tw} reduces the
problem of finding single, static, charged, spherically-symmetric
black-hole solutions of a generic 4-dimensional theory of gravity
coupled to a number of Abelian vectors $A^{\Lambda}{}_{\mu}$ and
scalars $\phi^{i}$ (without scalar potential) to the simpler problem
of finding solutions to a dynamical system whose dynamical variables
are just the metric function $U(\tau)$ and the scalar fields
$\phi^{i}(\tau)$; the evolution parameter $\tau$ corresponds to a
radial coordinate in the black hole spacetime metric.  This dramatic
simplification allowed the authors of Ref.~\cite{Ferrara:1997tw} to
derive the very important result, valid for the extremal black-hole
solutions of any of these theories including all the 4-dimensional
ungauged supergravity theories, relating the attractor values of the
scalars on the event horizon with the entropy through the so-called
black-hole potential. We will refer to this famous result as the
\textit{FGK theorem}.

Following these results, most of the work in this field has focused on
extremal black holes (supersymmetric and non-supersymmetric) since
they can be characterized, to a large extent, by the possible
attractors and the entropy which, in many supersymmetric theories with
large enough duality groups, can be determined by purely algebraic
methods.

The FGK formalism was not used for the explicit construction of the
extremal solutions, though. The dynamical system is simpler than the
original equations but still very non-linear and complicated. The
supersymmetric extremal solutions were constructed by methods based on
the study of the consistency conditions of the Killing spinor
equations. Even though the form of these solutions is known, showing
that they solve the equations of motion of the FGK formalism is not a
simple task. Non-supersymmetric extremal solutions have received a lot
of attention in the last few years: there are more of these than
supersymmetric ones and, furthermore, they have a richer structure. A
first-order formalism has been constructed for them starting from the
FGK dynamical system and a lot has been learned about the possible
attractors, entropies etc., see {\em e.g.\/}
Refs.~\cite{Miller:2006ay,Galli:2010mg}.  However, not many explicit
solutions have been constructed since the first-order equations are
not easy to integrate.

Non-extremal black-hole solutions have been left untouched by these
developments since the FGK theorem does not apply to them: one needs
to construct the explicit solution in order to compute the entropy,
the temperature and the dependence of the very important
\textit{non-extremality} parameter $r_{0}$ on the physical constants,
{\em i.e.\/} mass, electric and magnetic charges and the values of the
scalars at infinity. In Ref.~\cite{Galli:2011fq} a general ansatz for
non-extremal black holes of ungauged $N=2,d=4$ supergravity was
proposed and it was shown that using this ansatz the equations of
motion of the FGK formalism can be solved at least for some simple
theories\footnote{ A generalization of the FGK formalism for
  higher-dimensional theories as made in Ref.~\cite{Meessen:2011bd},
  where a similar ansatz was shown to work in a simple $N=2$, $d=5$
  supergravity theory.  }.  Non-extremal solutions interpolate between
different extremal solutions, supersymmetric and non-supersymmetric
alike, that can be recovered by taking the extremal limit. This
provides a new method for constructing the extremal non-supersymmetric
solutions.

The \textit{hyperbolic ansatz} proposed in Ref.~\cite{Galli:2011fq}
was based on the assumption that all the black-hole solutions of a
given theory have exactly the same expression in terms of some
functions $H^{M}(\tau)$, called {\em seed functions}. Different
solutions correspond to different profiles for the seed functions,
since they will satisfy different equations. For supersymmetric
solutions, the functions $H^{M}(\tau)$ will just be harmonic functions
(linear in the coordinate $\tau$). For non-extremal solutions,
Ref.~\cite{Galli:2011fq} proposed that the seed functions
$H^{M}(\tau)$ should be linear combinations of hyperbolic
functions. The hyperbolic ansatz was known to be valid in the few
non-extremal solutions known to the literature
\cite{Kallosh:1993yg,LozanoTellechea:1999my}. Furthermore, the
expression of the physical fields in terms of the $H^{M}(\tau)$ was
known to remain the same after the gauging of global symmetries
\cite{Huebscher:2007hj}.

The assumption that the black hole solutions have the same form in
terms of the seed functions was proven in the formulation of the H-FGK
formalism for $N=2,d=4$ supergravity theories, developed in
Refs.~\cite{Mohaupt:2011aa,Meessen:2011aa}: this formalism is obtained
from the standard FGK one by a change of variables, the new variables
being, precisely, the $H^{M}$s mentioned above\footnote{ This
  formulation is clearly related to the real formulation of local
  special geometry of Ref.~\cite{Ferrara:2006at}.  }.  The very
existence of the change of variables in all $N=2,d=4$ theories proves
the assumption. However, the new formulation has additional
advantages: since the new variables are, somehow, the ``right''
variables, finding new solutions and general results (attractor
theorems, first-order flow equations etc.) becomes much simpler
\cite{kn:GOPS}\footnote{ There is also an H-FGK formulation for black
  holes and black strings of $N=2,d=5$ supergravity
  \cite{Mohaupt:2009iq,Mohaupt:2010fk,Meessen:2011aa}. The derivation
  of the attractor theorem, first-order flow equations etc.~has been
  done in Ref.~\cite{Meessen:2012su}.  }.  In particular, it is
extremely easy to prove that the supersymmetric extremal black-hole
solutions with harmonic $H^{M}$s are solutions of the equations of
motion; the situation w.r.t.~extremal non-supersymmetric black hole
solutions is more complicated.

There are, however, some loose ends in these developments: in
Ref.~\cite{Lopes Cardoso:2007ky,Gimon:2009gk} an extremal
non-supersymmetric solution for cubic models was constructed
in which one of the $H^{M}(\tau)$s, rather than being harmonic, has been shown in Ref.~\cite{Galli:2010mg}
 to be the inverse of a harmonic function. Ratios of harmonic functions 
have been later on discussed and confirmed in Ref.~\cite{Bossard:2012xs,kn:GGP}. On the other hand, the general
study performed in \cite{kn:GOPS} suggests that in extremal black
holes, supersymmetric or not, all the $H^{M}$s should be
harmonic\footnote{ Observe that the hyperbolic ansatz always gives
  harmonic functions in the extremal limit.}.  Furthermore, the
hyperbolic ansatz is used together with a simplifying constraint on
the variables $H^{M}$ which arises quite naturally in the
supersymmetric case \cite{Bellorin:2006xr}, but which has no
justification in the non-supersymmetric cases, both extremal an
non-extremal. The non-harmonic solutions of Refs.~\cite{Lopes
  Cardoso:2007ky,Gimon:2009gk,Galli:2010mg,Bossard:2012xs,kn:GGP} do
not satisfy said constraint.

In this paper we take a first step towards the clarification of the
situation by showing how the description of a solution in terms of the
variables $H^{M}$ is not unique. We are going to show the existence of
a gauge symmetry in the 4-dimensional H-FGK formalism that acts on the
variables $H^{M}$ in a highly non-trivial and non-linear way but
preserves the physical fields of the black-hole solution: the metric
function $U(\tau)$ and the complex scalar fields $Z^{i}(\tau)$. This
symmetry does not preserve the above-mentioned constraint and, as we
are going to see, it can relate a configuration of the $H^{M}$s that
does not satisfy it to another configuration that does: both
configurations, however, describe the same physical black-hole
solution.  Whether the transformed $H^{M}$ that do satisfy the
constraint are harmonic is more difficult to prove in general and we
will study this problem in another publication \cite{kn:GOP}.

An interesting aspect of the gauge symmetry that we have discovered is
that it is based on a generalization of the Freudenthal duality
transformation discovered in Ref.~\cite{Borsten:2009zy} and
generalized in the context of $\mathcal{N}=8,d=4$ supergravity and
generalized to $\mathcal{N} \geq 2, d=4$ supergravities in
Ref.~\cite{Ferrara:2011gv}. The original Freudenthal transformation is
a discrete transformation that acts on the symplectic vector of
magnetic and electric charges of a given theory\footnote{The
  transformation depends on the particular theory under
  consideration.}  but one can define the same action on any other
symplectic vector of the same theory and, in particular on the
variables $H^{M}$. As we will show, the discrete transformations are a
particular case of a continuous local symmetry of the H-FGK.

We start by reviewing in depth the H-FGK formalism for $N=2,d=4$
theories in section (\ref{sec:HFGK}).  In section (\ref{sec:discrete})
we discuss the discrete Freudenthal transformations and in section
(\ref{sec:local}) we show that the HFGK action has a Freudenthal gauge
symmetry.  In section (\ref{sec:unconventional}) we discuss the
interplay of the Freudenthal gauge symmetry with the constraint,
identifying the latter as a gauge fixing condition. Finally, in
Sec.~(\ref{sec:conclusions}) we present our conclusions and discuss,
briefly, the implications of the local Freudenthal symmetry for the
extremal solutions.

\section{The H-FGK formalism for $\mathcal{N}=2$, $d=4$ supergravity
  revisited}
\label{sec:HFGK}

The action of all ungauged $N=2,d=4$ supergravity theories coupled to
$n$ vector multiplets takes the form\footnote{We will follow the
  notation and conventions of Ref.~\cite{Meessen:2011aa}.}

\begin{equation}
\label{eq:generalactionN2d4}
\begin{array}{rcl}
I[g_{\mu\nu},A^{\Lambda}{}_{\mu},Z^{i}]
& = &
{\displaystyle\int} d^{4}x \sqrt{|g|}
\left\{
R +2\mathcal{G}_{ij^{*}}\partial_{\mu}Z^{i}\partial^{\mu}Z^{*\, j^{*}}
+2 \Im\mathfrak{m} \mathcal{N}_{\Lambda\Sigma}
F^{\Lambda}{}_{\mu\nu}F^{\Sigma\, \mu\nu}
\right.
\\
& & \\
& & 
\left.
-2 \Re\mathfrak{e} \mathcal{N}_{\Lambda\Sigma}
F^{\Lambda}{}_{\mu\nu}\star F^{\Sigma\, \mu\nu}
\right\}\, ,   
\end{array}
\end{equation}

\noindent
where $i,j=1,\ldots, n$ and $\Lambda,\Sigma=0,1,\ldots,n$. The
scalar-dependent K\"ahler metric $\mathcal{G}_{ij^{*}}$ and period
matrix $\mathcal{N}_{\Lambda\Sigma}$ are related by supersymmetry and
can be derived, in general, from a holomorphic prepotential function
$\mathcal{F}(\mathcal{X})$ homogeneous of degree 2 in the coordinates
$\mathcal{X}^{\Lambda}$ or, equivalently, from a canonically
normalized, covariantly holomorphic symplectic section
$(\mathcal{V}^{M}) = \bigl(\begin{smallmatrix} \mathcal{L}^{\Lambda} \\
  \mathcal{M}_{\Lambda}
\end{smallmatrix} \bigr)$. Here $M,N,\ldots$ are $(2n+2)$-dimensional
symplectic indices and  we use the
symplectic metric $\left(\Omega_{MN} \right) \equiv \bigl(\begin{smallmatrix}
0 & \mathbbm{1}\\ -\mathbbm{1} & 0
\end{smallmatrix} \bigr)$ and $\Omega^{MP}\Omega_{NP}= \delta^{M}{}_{N}$ to
lower and rise the symplectic indices according to the convention

\begin{equation}
  \mathcal{V}_{M} = \Omega_{MN}\mathcal{V}^{N}\, ,
  \hspace{1cm}  
  \mathcal{V}^{M} = \mathcal{V}_{N}\Omega^{NM}\, .
\end{equation}

The metrics of all the single, static, 4-dimensional black-hole
solutions to these theories can be put in the form

\begin{equation}
\label{eq:generalbhmetric}
\begin{array}{rcl}
ds^{2} 
& = & 
e^{2U} dt^{2} - e^{-2U} \gamma_{\underline{m}\underline{n}}
dx^{\underline{m}}dx^{\underline{n}}\, ,  \\
& & \\
\gamma_{\underline{m}\underline{n}}
dx^{\underline{m}}dx^{\underline{n}}
& = & 
{\displaystyle\frac{r_{0}^{4}}{\sinh^{4} r_{0}\tau}}d\tau^{2} 
+
{\displaystyle\frac{r_{0} ^{2}}{\sinh^{2}r_{0}\tau}}d\Omega^{2}_{(2)}\, ,\\
\end{array}
\end{equation}

\noindent
where $r_{0}$ is the so-called {\em non-extremality parameter} and
$U(\tau)$ the metric function that characterizes a particular
solution\footnote{ More information about this metric can be found in
  Ref.~\cite{Galli:2011fq}.  }.  Assuming that all the fields are
static and spherically symmetric, so that they only depend on the
radial coordinate $\tau$, the action (\ref{eq:generalactionN2d4})
reduces to the FGK effective action \cite{Ferrara:1997tw}

\begin{equation}
\label{eq:effectiveaction}
I_{\text{FGK}}[U,Z^{i}] = \int d\tau \left\{ 
(\dot{U})^{2}  
+\mathcal{G}_{ij^{*}}\dot{Z}^{i}  \dot{Z}^{*\, j^{*}}  
-e^{2U}V_{\rm bh}(Z,Z^{*},\mathcal{Q})
\right\}\, ,  
\end{equation}

\noindent
which has to be supplemented by the Hamiltonian constraint

\begin{equation}
\label{eq:HamConstr}
(\dot{U})^{2}  
+\mathcal{G}_{ij^{*}}\dot{Z}^{i}  \dot{Z}^{*\, j^{*}}  
+e^{2U}V_{\rm bh}(Z,Z^{*},\mathcal{Q})
= 
r_{0}^{2}\, .  
\end{equation}

\noindent
In the above formulae $V_{\rm bh}(Z,Z^{*},\mathcal{Q})$ is the so-called
\textit{black-hole potential} and is given by

\begin{equation}
\label{eq:BHpot}
-V_{\rm bh}(Z,Z^{*},\mathcal{Q})
\; =\; 
-\tfrac{1}{2}
\mathcal{M}_{MN}(\mathcal{N})\mathcal{Q}^{M}\mathcal{Q}^{N}\, ; 
\end{equation}

\noindent
$\mathcal{Q}^{M}$ is the $(2n+2)$-dimensional symplectic vector of
electric $q$ and magnetic $p$ charges $(\mathcal{Q}^{M}) =
\bigl(\begin{smallmatrix} p^{\Lambda} \\ q_{\Lambda} \end{smallmatrix}
\bigr)$ and $\mathcal{M}_{MN}(\mathcal{N})$ is the symmetric,
symplectic matrix defined by

\begin{equation}
\label{eq:MofN}
\left(\mathcal{M}_{MN}(\mathcal{N}) \right)
\equiv 
\left(
\begin{array}{cc}
I+RI^{-1}R  & \,\,\,\, -RI^{-1} \\
& \\
-I^{-1}R & I^{-1} \\   
\end{array}
\right)\, ,
\hspace{1cm}
R \equiv \Re\mathfrak{e}\, \mathcal{N}\, ,
\hspace{.5cm}
I \equiv \Im\mathfrak{m}\, \mathcal{N}\, .
\end{equation}

\noindent
Observe that since there is no explicit $\tau$ dependence in the
effective action (\ref{eq:effectiveaction}), the corresponding
Hamiltonian must take a constant value: the Hamiltonian constraint
(\ref{eq:HamConstr}) fixes this {\em a priori} unconstrained value to
be $r_{0}^{2}$.

The change of variables that brings us to the H-FGK formalism is
inspired in the general form of the timelike supersymmetric solutions
of these theories obtained by analyzing the consistency of the Killing
spinor equations (see {\em e.g.\/} Ref.~\cite{Meessen:2006tu}): given
an $\mathcal{N}=2$, $d=4$ theory with canonical symplectic section
$\mathcal{V}^{M}$, introducing a complex variable $X$ with the same
K\"ahler weight as $\mathcal{V}^{M}$, we can define the real
K\"ahler-neutral symplectic vectors

\begin{equation}
\label{eq:RandIdef}
\mathcal{R}^{M}  \; \equiv\; \Re\mathfrak{e}\left( \mathcal{V}^{M}/X\right) \; ,
\hspace{1cm}
\mathcal{I}^{M} \; \equiv\; \Im\mathfrak{m}\left(  \mathcal{V}^{M}/X\right) \; .
\end{equation}

\noindent
The components $\mathcal{R}^{M}$ can be expressed in terms of the
$\mathcal{I}^{M}$ by solving a set of algebraic equations commonly
called the stabilization equations \cite{Ferrara:1996dd} (although
this name is used with a different meaning in part of the literature),
but to which we shall refer henceforth, for reasons that will become
clear in the following and to avoid confusion, as the {\em Freudenthal
  duality equations}.  The functions $\mathcal{R}^{M}(\mathcal{I})$
are characteristic of each theory, but they are always homogeneous of
first degree in the $\mathcal{I}^{M}$.

Given the fact that, in supersymmetric solutions, the
$\mathcal{I}^{M}$ are harmonic functions, it is customary to relabel
these variables as

\begin{equation}
H^{M} \;\equiv\; \mathcal{I}^{M}\, ,
\hspace{1.5cm}
\tilde{H}^{M} \;\equiv\; \mathcal{R}^{M}\, .
\end{equation}

Given those functions we can define the \textit{Hesse potential}
$\mathsf{W}(H)$ \cite{Bates:2003vx,Mohaupt:2011aa,Meessen:2011aa}
 
\begin{equation}
\mathsf{W}(H) 
\;\equiv\; 
\langle\, \tilde{H}\mid H \,  \rangle 
\equiv 
\tilde{H}_{M} H^{M}\, ,  
\end{equation}

\noindent
which is homogeneous of second degree in $H^{M}$.  The relation
between $\tilde{H}^{M}$ and $H^{M}$ can be inverted and the Hesse
potential can also be written as $\mathsf{W}(\tilde{H})$; from the
homogeneity of $\mathsf{W}$ one can deduce that

\begin{equation}
\label{eq:2}
\tilde{H}_{M} 
\; =\; 
\tfrac{1}{2} \frac{\partial\mathsf{W}}{\partial H^{M}}
\;\equiv\; 
\tfrac{1}{2} \partial_{M}\mathsf{W}\, ,
\hspace{1cm}
  H^{M}
\; =\; 
\tfrac{1}{2} \frac{\partial\mathsf{W}}{\partial \tilde{H}_{M}} \; .
\end{equation}

Of special importance to the H-FGK formalism is the symmetric
symplectic matrix $\mathcal{M}_{MN}(\mathcal{F})$ which is obtained by
replacing in the expression (\ref{eq:MofN}) the period matrix
$\mathcal{N}_{\Lambda\Sigma}$ by

\begin{equation}
\mathcal{F}_{\Lambda\Sigma}
\equiv 
\frac{\partial^{2}\mathcal{F}(\mathcal{X})}{\partial \mathcal{X}^{\Lambda}\partial\mathcal{X}^{\Sigma}}\, ,
\end{equation}

\noindent
where $\mathcal{F}(\mathcal{X})$ is the prepotential of the theory;
the relation between them can be seen to be

\begin{equation}
\label{eq:3}
\mathcal{M}_{MN}(\mathcal{F}) 
\; =\; 
-\mathcal{M}_{MN}(\mathcal{N}) \ -2\mathsf{W}^{-1}\ ( H_{M}H_{N}
 \ +\tilde{H}_{M}\tilde{H}_{N})\, .
\end{equation}

From the fundamental properties of the matrix
$\mathcal{M}(\mathcal{F})$, namely

\begin{equation}
\label{eq:RM}
\begin{array}{cccccc}
\tilde{H}_{M} 
& = & 
-\mathcal{M}_{MN}(\mathcal{F}) H^{N}\, , \hspace{1cm} &    
d\tilde{H}_{M} 
& = & 
-\mathcal{M}_{MN}(\mathcal{F}) dH^{N} \; ,\\
& & & & & \\
H_{M} 
& = & 
\mathcal{M}_{MN}(\mathcal{F}) \tilde{H}^{N}\, ,  & 
dH_{M} 
& = & 
\mathcal{M}_{MN}(\mathcal{F}) d\tilde{H}^{N} \; ,
  \end{array}
\end{equation}

\noindent
one can infer that

\begin{equation}
\label{eq:HessianversusF}
\mathcal{M}_{MN}(\mathcal{F}) 
\; =\; 
-\tfrac{1}{2} 
\frac{\partial^{2}\mathsf{W}}{\partial H^{M}\partial H^{N}}
\; =\; 
\tfrac{1}{2} 
\frac{\partial^{2}\mathsf{W}}{\partial \tilde{H}^{M}\partial \tilde{H}^{N}} \; ,
\end{equation}

\noindent
this equation can be rewritten using eqs.~(\ref{eq:2}) as

\begin{equation}
\label{eq:4}
\frac{\partial\tilde{H}_{N}}{\partial H^{M}} 
\; =\; 
\Omega_{MP}\Omega_{NQ}\ \frac{\partial H^{Q}}{\partial \tilde{H}_{P}} \; ,
\end{equation}

\noindent
which is equivalent to saying that $\mathcal{M}$ is a symplectic
matrix.

Eq.~(\ref{eq:HessianversusF}) tells us that the Hesse potential
$\mathsf{W}$ is closely related to the prepotential and is to be
considered a \textit{real prepotential}.

Observe that the above discovered Hessianity implies that
$\partial_{P}\mathcal{M}_{MN}(\mathcal{F})
=\partial_{(P}\mathcal{M}_{MN)}(\mathcal{F})$, whereas the homogeneity
implies

\begin{equation}
  \label{eq:6}
    0 \; =\; H^{P}\partial_{P}\mathcal{M}_{MN}(\mathcal{F}) 
       \; =\; \tilde{H}^{P}\partial_{P}\mathcal{M}_{MN}(\mathcal{F}) \; .
\end{equation}

Now, using general properties of Special Geometry and the above
properties one can rewrite the effective action
(\ref{eq:effectiveaction}) and Hamiltonian constraint
(\ref{eq:HamConstr}) entirely in terms of the new variables $H^{M}$
\cite{Meessen:2011aa}:

\begin{eqnarray}
\label{eq:effectiveaction2}
-I_{\text{H-FGK}}[H] 
& = & 
\int d\tau 
\left\{ 
\tfrac{1}{2}g_{MN}
\dot{H}^{M}\dot{H}^{N}
-V
\right\}\, ,
\\
& & \nonumber \\
\label{eq:hamiltonianconstraint}
r_{0}^{2} & =& \tfrac{1}{2}g_{MN}
\dot{H}^{M}\dot{H}^{N}
\ +\ V  \; ,
\end{eqnarray}

\noindent
where we have defined the $H$-dependent metric

\begin{equation}
g_{MN}\; \equiv\; 
\partial_{M}\partial_{N}\log{\mathsf{W}}
-2\frac{H_{M}H_{N}}{\mathsf{W}^{2}}
\; =\;
\frac{\partial_{M}\partial_{N}\mathsf{W}}{\mathsf{W}}
-2\frac{H_{M}H_{N}}{\mathsf{W}^{2}}
-4\frac{\tilde{H}_{M}\tilde{H}_{N}}{\mathsf{W}^{2}}\, ,
\end{equation}

\noindent
and the potential 

\begin{equation}
\label{eq:potential}
V(H) 
\ =\ 
\left\{
-\tfrac{1}{4}\partial_{M}\partial_{N}\log{\mathsf{W}}
+\frac{H_{M}H_{N}}{\mathsf{W}^{2}}
\right\}\mathcal{Q}^{M}\mathcal{Q}^{N}
=
\left\{
-\tfrac{1}{4}g_{MN}
+\tfrac{1}{2}\frac{H_{M}H_{N}}{\mathsf{W}^{2}}
\right\}\mathcal{Q}^{M}\mathcal{Q}^{N}\, .
\end{equation}

\noindent
The relation of this potential to the black-hole potential
(\ref{eq:BHpot}) is given by

\begin{equation}
V_{\rm bh} \; =\; \mathsf{W}\ V\, .  
\end{equation}


\section{Discrete Freudenthal transformations}
\label{sec:discrete}

The relation between the tilded and untilded variables can be
understood as a duality transformation $H^{M} \rightarrow
\tilde{H}^{M}$ which can be iterated if we define
$\tilde{\tilde{H}}^{M} \equiv \tilde{H}^{M}(\tilde{H})$. Using the
properties in Eqs.~(\ref{eq:2}--\ref{eq:6}), we find that this duality
is an anti-involution, {\em e.g.\/}

\begin{equation}
\label{eq:antiinvolution}
\tilde{\tilde{H}}^{M} \; =\; -H^{M}\, .  
\end{equation}

It is not difficult to see that the duality transformation is just the
generalization to $\mathcal{N}=2$, $d=4$ supergravity theories made in
Ref.~\cite{Ferrara:2011gv} of the \textit{Freudenthal duality}
introduced in Ref.~\cite{Borsten:2009zy} in the context of
$\mathcal{N}=8$, $d=4$ supergravity. The same operation can be
performed on any symplectic vector of a given theory and, in
particular, on the charge vector $\mathcal{Q}$.

In Ref.~\cite{Ferrara:2011gv} it was shown that the entropy and the
critical points of the black-hole potential are invariant under
Freudenthal duality. We will recover this result later as a particular
case of the invariance of the H-FGK system under \textit{local
  Freudenthal rotations}.

The variables we have just defined are related to the physical
variables of the FGK formalism $U$, $Z^{i}$ by
\cite{Meessen:2011aa}\footnote{The expression for the scalars is not
  unique (only up to reparametrizations). The expression we give is,
  however, convenient and simple.}

\begin{equation}
\label{eq:changeofvariables}
e^{-2U}\;\equiv\; \mathsf{W}(H)= \tilde{H}_{M}H^{M}\, ,  
\hspace{1.5cm}
Z^{i} \;\equiv\; \frac{\tilde{H}^{i}+iH^{i}}{\tilde{H}^{0} +iH^{0}}\, .  
\end{equation}

\noindent
We can immediately see that the physical variables are invariant under
the above Freudenthal duality transformations, {\em i.e.\/}

\begin{equation}
e^{-2U}(\tilde{H}) = e^{-2U}(H)\, ,
\hspace{1.5cm}  
Z^{i}(\tilde{H}) = Z^{i}(H)\, ,
\end{equation}

It is interesting to study how the central charge changes under
Freudenthal duality: first, we rewrite the central charge, whose
definition is $\mathcal{Z}(\phi,\mathcal{Q})\equiv
\langle\,\mathcal{V} \mid \mathcal{Q} \,\rangle$ in the form

\begin{equation}
\mathcal{Z}(\phi,\mathcal{Q}) = \frac{e^{i\alpha}}{\sqrt{2\mathsf{W}(H)}} 
(\tilde{H}_{M} +i H_{M})\mathcal{Q}^{M}\, ,  
\end{equation}

\noindent
where $e^{i\alpha}$ is the phase of $X$ and satisfies the equation
\cite{Meessen:2006tu}

\begin{equation}
\label{eq:alphaequation}
\dot{\alpha} \; =\; \mathsf{W}^{-1}\ \dot{H}^{M}H_{M} \; -\; \mathcal{Q}_{\star}\, ,
\end{equation}

\noindent
where $\mathcal{Q}_{\star}$ is the pullback of the K\"ahler connection 1-form

\begin{equation}
\mathcal{Q}_{\star} 
= 
\tfrac{1}{2i}\dot{Z}^{i}\partial_{i}\mathcal{K} +\mathrm{c.c.}  
\end{equation}

Under discrete Freudenthal duality transformations, $\mathsf{W}(H)$,
the scalars and the K\"ahler potential are invariant. $\alpha$ is also
invariant and

\begin{equation}
(\tilde{H}_{M} +i H_{M})^{\prime} = -i (\tilde{H}_{M} +i H_{M})\, ,   
\end{equation}

\noindent
which implies that 

\begin{equation}
\mathcal{Z}^{\prime}(\phi,\mathcal{Q}) = 
-i\mathcal{Z}(\phi,\mathcal{Q})\, , 
\end{equation}

\noindent
but its absolute value will remain invariant. 

Observe that when these Freudenthal transformations are non-linear
(which is the general case), if we transform a supersymmetric
solution, which must have harmonic $H^{M}$s of the form

\begin{equation}
H^{M}=A^{M} -\tfrac{1}{\sqrt{2}}\mathcal{Q}^{M} \tau\, ,  
\end{equation}

\noindent
we will obtain non-harmonic $H^{M}$ and the transformed solution
couldn't possibly be supersymmetric. We must remember, however, that
all the physical fields are invariant, whence their supersymmetry
properties must also remain invariant. This implies that the variables
$H^{M}$ cannot immediately be identified with those appearing in the
analysis of the Killing spinor equations: this is possible only up to
discrete Freudenthal transformations.

The near-horizon limit of the transformed $H^{M}$s is dominated by the
Freudenthal dual of the charges $\mathcal{Q}^{M}$, defined in
Refs.~\cite{Borsten:2009zy,Ferrara:2011gv}, namely

\begin{equation}
\label{eq:1}
\tilde{\mathcal{Q}}^{M}
\; \equiv\; 
-\tfrac{1}{2}\ \Omega^{MN}\ 
\frac{\partial \mathsf{W}(\mathcal{Q})}{\partial \mathcal{Q}^{N}} \; .
\end{equation}


\section{Local Freudenthal rotations}
\label{sec:local}

In the change of variables taking us to the H-FGK formalism, we have
gone from a formulation based on $2n+1$ real variables, namely $U$ and
the $Z^{i}$, to one which is based on $2n+2$ variables, whence we
obtained an over-complete formulation.  This suggests that there should
be a local symmetry in the H-FGK formalism allowing the elimination of
one of its degrees of freedom. The variables $H^{M}$, on the other
hand, transform linearly under the duality group (embedded in
$\mathrm{Sp}(n+1;\mathbb{R})$), as follows from its definition.

The looked-for gauge symmetry can be found by observing that the
metric $g_{MN}$ is singular: using the properties
(\ref{eq:RM}--\ref{eq:6}) it is easy to show that it always admits an
eigenvector with zero eigenvalue, namely\footnote{For the sake of
  completeness we also quote the relation
\begin{equation}
g_{MN}H^{N} =-2 \tilde{H}_{M}/\mathsf{W}
\,\,\,\,\,
\Rightarrow
\,\,\,\,\, 
g_{MN}H^{M}H^{N} =-2\, .
\end{equation}
}:

\begin{equation}
\tilde{H}^{M}g_{MN}=0\, .  
\end{equation}

The equations of motion in the H-FGK formalism are

\begin{equation}
\label{eq:equationsofmotion}
\frac{\delta I_{\text{H-FGK}} }{\delta H^{M}}
=
g_{MN}\ \ddot{H}^{N} \, +\, \left[ PQ , M\right]\ \dot{H}^{P}\dot{H}^{Q} 
\, +\, \partial_{M}V=0\, ,  
\end{equation}

\noindent
where, as $g_{MN}$ is not invertible, we have used the Christoffel symbol of
the first kind, {\em i.e.\/}

\begin{equation}
 \textstyle{[PQ, M]}\, \equiv\, \partial_{(P}g_{Q)M}
  \ -\ \tfrac{1}{2} \partial_{M}g_{PQ}\, . 
\end{equation}

\noindent
Using the properties (\ref{eq:RM}--\ref{eq:6}) it is not difficult to show
that

\begin{equation}
\label{eq:gaugeidentity}
\left.
  \begin{array}{ccc}
    [PQ,M]\ \tilde{H}^{M} & =& 0\\
      & & \\
    \tilde{H}^{M}\partial_{M}V & =& 0
  \end{array}
\right\} \;\;\;\;\xrightarrow{\hspace{.7cm}\mbox{so that}\hspace{.7cm}}\hspace{1.2cm}
\tilde{H}^{M}\frac{\delta I_{\text{H-FGK}} }{\delta H^{M}} =0\, . 
\end{equation}

This is a constraint that relates the equations of motion of the H-FGK
formalism. This kind of constraints arises in systems with gauge
symmetries, as consequence of Noether's second theorem and is a
\textit{gauge identity}.  Indeed, multiplying the constraint by an
arbitrary infinitesimal function $f(\tau)$ and integrating over $\tau$
we find that Eq.~(\ref{eq:gaugeidentity}) implies

\begin{equation}
\delta_{f} I_{\text{H-FGK}} = \int d\tau \delta_{f}H^{M}\frac{\delta
  I_{\text{H-FGK}} }{\delta H^{M}} =0\, ,  
\end{equation}

\noindent
where we have defined the local infinitesimal transformations

\begin{equation}
 \delta_{f}H^{M} \equiv   f(\tau) \tilde{H}^{M}\, .
\end{equation}

As one can expect from a gauge invariance, this transformation leaves
invariant the physical variables of the FGK formalism $U$, $Z^{i}$. To
check it, it is enough to use

\begin{equation}
\label{eq:infiFreud}
 \delta_{f}\tilde{H}^{M} \equiv   -f(\tau) H^{M}\, ,
\end{equation}

\noindent
which follows from Eq.~(\ref{eq:antiinvolution}) and
Eqs.~(\ref{eq:changeofvariables}).

The finite gauge transformations can be obtained by exponentiating the
infinitesimal ones:

\begin{equation}
\delta_{f}H^{M} \equiv f(\tau)\pounds_{K} H^{M}
\,\,\,\,\,\,
\longrightarrow
\,\,\,\,\,\,
H^{\prime\, M} = e^{f(\tau)\pounds_{K}} H^{M}
\;\;\mbox{where}\;\;  K^{M}(H) = \tilde{H}^{M}\, .
\end{equation}

\noindent
It is not difficult to see that the finite transformations are 

\begin{equation}
\left\{
\begin{array}{rcl}
H^{\prime\, M} 
& = &
\cos{f}\ H^{M} -\sin{f}\ \Omega^{MN}\tilde{H}_{N}\, ,   
\\
& &  \\
\tilde{H}^{\prime}_{M} 
& = &
-\sin{f}\ \Omega_{MN}H^{N}+\cos{f}\ \tilde{H}_{M}\, .  
\end{array}
\right.
\end{equation}

\noindent
By defining the complex variables $\mathcal{H}^{M}\equiv \tilde{H}^{M}
+i H^{M}$ we can write the transformation as

\begin{equation}
\mathcal{H}^{\prime\, M}
=
e^{if(\tau)}
\mathcal{H}^{M}\, .
\end{equation}

\noindent
Using this form of the transformation and expressing the scalars and the
metric function in the forms

\begin{equation}
e^{-2U}
=
\mathsf{W}(H)
=
\tfrac{i}{2}
\mathcal{H}_{M}
\mathcal{H}^{*\, M}
\, ,
\hspace{1cm}
Z^{i} \equiv \mathcal{H}^{i}/\mathcal{H}^{0}\, ,
\end{equation}

\noindent
the invariance of the physical fields under this gauge symmetry is paramount.

A direct proof of the invariance of the H-FGK effective action is also
desirable: the invariance of the kinetic term, {\em i.e.\/}
$\tfrac{1}{2}g_{MN}\dot{H}^{M}\dot{H}^{N}$, follows from the
identities

\begin{equation}
(\tilde{H}_{M}\dot{H}^{M})^{\prime} = \tilde{H}_{M}\dot{H}^{M}\, ,
\hspace{1cm}
\dot{\tilde{H}}{}^{M}\mathcal{M}_{MN}(\mathcal{F}) 
=
\dot{H}_{N}\, ,
\hspace{1cm}
\dot{H}^{M}\mathcal{M}_{MN}(\mathcal{F}) 
=
-\dot{\tilde{H}}_{N}\, ,
\end{equation}

\noindent
which can be derived from Eqs.~(\ref{eq:RM}). The invariance of the potential
$V(H)$ follows from Eq.~(\ref{eq:6}).

The existence of this symmetry does not help in solving the equations
of motion as the Noether charge associated to the invariance under the
global Freudenthal rotations vanishes identically:

\begin{equation}
\mathsf{Q} = \delta_{f} H^{M}\frac{\partial L}{\partial H^{M}} 
\sim f \tilde{H}^{M}g_{MN}\dot{H}^{N}=0\, .
\end{equation}

We have already said that the origin of this gauge symmetry is the
introduction of one additional degree of freedom in the passage from
the FGK to the H-FGK formalism.  Had the original FGK formulation
contained the full complex variable $X= e^{U+i\alpha}$ instead of just
$U$, the change of variables would, actually, have been much simpler;
alas, the phase $\alpha$ is completely absent from the FGK effective
action. The local Freudenthal symmetry is associated to this absence,
which allows to change $\alpha$ arbitrarily leaving everything else
invariant. Indeed, from Eq.~(\ref{eq:alphaequation}) that defines
$\alpha$, we can easily see that

\begin{equation}
\delta_{f} \dot{\alpha} = -\dot{f}\, .
\end{equation}

On the other hand, the Freudenthal gauge symmetry can be made manifest as
follows: first, observe that the metric 

\begin{equation}
G_{MN}(H) 
\equiv
\partial_{M}\partial_{N}\log{\mathsf{W}} -2(1+\varepsilon)
\frac{H_{M}H_{N}}{\mathsf{W}}\, ,
\hspace{1cm}
\varepsilon=\pm 1\,   
\end{equation}

\noindent
always admits $K^{M}(H)=\tilde{H}^{M}$ as a Killing vector. Then, consider the
action

\begin{equation}
\label{eq:effectiveactionung}
-I_{\text{ungauged}}[H] 
= 
\int d\tau 
\left\{ 
\tfrac{1}{2}G_{MN}\dot{H}^{M} \dot{H}^{N}
-V
\right\}\, ,
\end{equation}

\noindent
which has a global Freudenthal symmetry generated by $\delta H^{M}= f
\tilde{H}^{M}$ with $\dot{f}=0$.  To gauge the Freudenthal symmetry,
we just have to replace in this action the derivatives with respect to
$\tau$ by the covariant derivatives

\begin{equation}
  \begin{array}{rcl}
\dot{H}^{M} 
 & \rightarrow &
\mathfrak{D}H^{M} \equiv \dot{H}^{M} +A \tilde{H}^{M}\, , \\
& & \\
\dot{\tilde{H}}^{M} 
 & \rightarrow &
\mathfrak{D}\tilde{H}^{M} \equiv \dot{\tilde{H}}^{M} -A H^{M}\, , \\
\end{array}
\end{equation}

\noindent
which transform covariantly under the infinitesimal transformations
Eq.~(\ref{eq:infiFreud})

\begin{equation}
  \begin{array}{rcl}
\delta_{f} \mathfrak{D} H^{M} 
& = & 
f \mathfrak{D} \tilde{H}^{M}\, ,
\\
& & \\    
\delta_{f} \mathfrak{D} \tilde{H}^{M} 
& = & 
-f \mathfrak{D} H^{M}\, ,
  \end{array}
\end{equation}

\noindent
if the 1-form $A$ transforms as

\begin{equation}
\delta_{f} A= -\dot{f}(\tau)\,.  
\end{equation}

\noindent
The 
action 

\begin{equation}
\label{eq:effectiveaction3}
-I_{\text{gauged}}[H,A] 
= 
\int d\tau 
\left\{ 
  \tfrac{1}{2}G_{MN}\mathfrak{D}H^{M} \mathfrak{D}H^{N}
  -V
\right\}\, ,
\end{equation}

\noindent
is manifestly invariant under local Freudenthal rotations and
equivalent to the effective H-FGK action
Eq.~(\ref{eq:effectiveaction2}) as one can see by integrating out the
auxiliary field $A$: its equation of motion is solved by

\begin{equation}
A =  \frac{H_{N}\dot{H}^{N}}{\mathsf{W}}\, ,  
\end{equation}

\noindent
and, upon this substitution

\begin{equation}
G_{MN}\mathfrak{D}H^{M} \mathfrak{D}H^{N}
= \left(G_{MN}+2\varepsilon\frac{H_{M}H_{N}}{\mathsf{W}} \right)
\dot{H}^{M}\dot{H}^{N}
=
g_{MN}\dot{H}^{M}\dot{H}^{N}\, .  
\end{equation}

The choice $\varepsilon=+1$, which leads to $G_{MN} =
2\mathsf{W}^{-1}\mathcal{M}_{MN}(\mathcal{N})$ is, perhaps, the most
natural since the same metric would then occur in the kinetic term and
in the potential. It follows that we can rewrite the effective
action Eq.~(\ref{eq:effectiveaction2}) and the Hamiltonian constraint
Eq.~(\ref{eq:hamiltonianconstraint}) in the suggestive form

\begin{eqnarray}
\label{eq:effectiveaction4}
I_{\text{H-FGK}}[H] 
& = & 
\int d\tau 
\left\{ 
V(H,\sqrt{2}\, \mathfrak{D}H)+V(H,\mathcal{Q})
\right\}\, ,
\\
& & \nonumber \\
\label{eq:hamiltonianconstraint2}
r_{0}^{2}
& = & 
V(H,\sqrt{2}\, \mathfrak{D}H)-V(H,\mathcal{Q})\, ,
\end{eqnarray}

\noindent
with 

\begin{equation}
\mathfrak{D}H^{M} 
=
\dot{H}^{M} + \frac{H_{N}\dot{H}^{N}}{\mathsf{W}}
\tilde{H}^{M}\, .   
\end{equation}

Finally, it is worth noting that this Freudenthal gauge theory is
unrelated to the one constructed in Ref.~\cite{Marrani:2012uu}.


\section{Unconventional solutions and Freudenthal gauge freedom}
\label{sec:unconventional}


If we contract the equations of motion (\ref{eq:equationsofmotion})
with $H^{P}$ and use the homogeneity properties of the different terms
and the Hamiltonian constraint Eq.~(\ref{eq:hamiltonianconstraint}),
we find a useful equation

\begin{equation}
\label{eq:Urewriten}
\tilde{H}_{M} \left(\ddot{H}^{M}  -r_{0}^{2}H^{M}\right) 
+
\frac{(\dot{H}^{M}H_{M})^{2}}{\mathsf{W}}  
\; =\;  
0\, ,
\end{equation}

\noindent
which corresponds to that of the variable $U$ in the FGK formulation.

In the supersymmetric (hence, extremal) case, the constraint

\begin{equation}
\label{eq:nonut}
\dot{H}^{M}H_{M} = 0\, , 
\end{equation}

\noindent
enforcing the absence of NUT charge must be satisfied, in agreement
with the assumption of staticity of the metric \cite{Bellorin:2006xr}.
Using this constraint the above equation takes the form

\begin{equation}
\label{eq:Urewriten2}
\tilde{H}_{M} \left(\ddot{H}^{M}
  -r_{0}^{2}H^{M}\right) 
= 0\, ,
\end{equation}

\noindent 
and can be solved in the extremal case by assuming that the $H^{M}$
are linear in $\tau$, whence they are harmonic, and in the
non-extremal case by assuming that the $H^{M}$ are linear combinations
of hyperbolic functions of $r_{0}\tau$ (the hyperbolic ansatz). The
solutions that one can get with these assumptions have been
intensively studied in Ref.~\cite{kn:GOPS}.

The constraint Eq.~(\ref{eq:nonut}) is not preserved by the local
Freudenthal symmetry: a small calculation gives

\begin{equation}
\delta_{f}(\dot{H}^{M}H_{M}) \; =\; -\dot{f} \mathsf{W}\, ,
\end{equation}

\noindent
which can be integrated straightforwardly to a finite rotation, namely

\begin{equation}
(\dot{H}^{M}H_{M})^{\prime} = -\dot{f} \mathsf{W} +\dot{H}^{M}H_{M}\, . 
\end{equation}

\noindent
This equation implies that given a configuration $H^{M}$ with
$\dot{H}^{M}H_{M}\neq 0$, we can find another configuration
$H^{\prime\, M}$ with $\dot{H}^{\prime\, M}H^{\prime}_{M}= 0$
\textit{describing exactly the same configuration of physical fields}
by performing a finite local Freudenthal transformation with a
parameter $f(\tau)$ satisfying

\begin{equation}
\dot{f} = \frac{\dot{H}^{M}H_{M}}{\mathsf{W}}\, .
\end{equation}

This shows that it is always possible to impose the constraint
Eq.~(\ref{eq:nonut}) without loss of generality because it can be
understood as just a good gauge-fixing condition.


\section{Conclusions}
\label{sec:conclusions}

The extremal static black-hole solutions of $\mathcal{N}=2,d=4$
supergravity constructed so far in the literature and written in terms
of the variables $H^{M}$ can be classified using two criteria: the
harmonicity of the $H^{M}$s and whether they satisfy the constraint
$H_{M}\dot{H}^{M}=0$ or not. Out of the four possible cases,
represented in table (\ref{tab:1}), the equation of motion
Eq.~(\ref{eq:Urewriten}) excludes the one corresponding to the upper
right corner. The upper left corner corresponds to the supersymmetric
black-hole solutions and, as shown in Ref.~\cite{Galli:2011fq}, also
to some non-BPS solutions as well. The lower-right corner corresponds
to the extremal non-BPS solutions discovered in Refs.~\cite{Lopes
  Cardoso:2007ky,Gimon:2009gk,Galli:2010mg,Bossard:2012xs,kn:GGP} and
the lower-left corner does not correspond to any known solution.

In this paper we have shown that the representation of the solutions
in terms of these variables is non-unique due to the presence of the
local Freudenthal invariance. Furthermore, we have shown that this
symmetry can be used to transform all the solutions in the lower-right
corner to solutions in the left column. It is not yet clear whether
they will be transformed into solutions in the upper or lower row
although preliminary results in simple examples suggest that,
typically, they will transformed into solutions in the lower-left
corner. The form of the $H^{M}$s in this class is probably quite
complicated as they must satisfy the equation

\begin{equation}
\tilde{H}_{M}\ddot{H}^{M} =0\, ,  
\end{equation}

\noindent
and, at the same time, $\ddot{H}^{M}\neq 0$. Furthermore, solutions of
this kind must be possible only in very special cases and only in some
theories, as it happens for the solutions in the lower-right corner.
Clearly, more work is needed to arrive at a complete understanding of
the situation and to chart the space of extremal black-hole solutions
of these theories. The non-extremal case is even more
challenging. Work in these directions is in progress \cite{kn:GOP}.

\begin{table}
\centering
\begin{tabular}{||l|c|c||}
\hline\hline
& & \\
& $H_{M}\dot{H}^{M}=0$ & $H_{M}\dot{H}^{M}\neq 0$ \\
\hline
& &  \\
$\ddot{H}^{M}=0$ & BPS and some non-BPS & no solutions \\
\hline
& & \\
$\ddot{H}^{M} \neq 0$ &  & some non-BPS \\
\hline\hline  
\end{tabular}
\caption{Classification of the extremal static black-hole 
solutions of  $\mathcal{N}=2,d=4$ supergravity according to their 
representation in terms of  the variables $H^{M}$. It must be taken 
into account that they satisfy  Eq.~(\ref{eq:Urewriten}) with $r_{0}=0$.}
\label{tab:1}
\end{table}


\section*{Acknowledgments}

PG and PM would like to thank the Instituto de F\'{\i}sica Te\'orica
UAM/CSIC its hospitality.  This work has been supported in part by the
Spanish Ministry of Science and Education grant FPA2009-07692, the
Princip\'au d'Asturies grant IB09-069, the Comunidad de Madrid grant
HEPHACOS S2009ESP-1473 and the Spanish Consolider-Ingenio 2010 program
CPAN CSD2007-00042. The work of PG has been supported in part by grants
FIS2008-06078-C03-02 and FIS2011-29813-C02-02 of Ministerio de Ciencia
e Innovaci\'on (Spain) and ACOMP/2010/213 from Generalitat
Valenciana. The work of PM has been supported by the Ram\'on y Cajal
fellowship RYC-2009-05014. TO wishes to thank M.M.~Fern\'andez for her
perduring support.

\appendix


%
%
\end{document}